\documentclass[%
 reprint,
superscriptaddress,
 amsmath,amssymb,
 aps,
]{revtex4-2}
\pdfoutput=1 
\usepackage{graphicx}
\usepackage{dcolumn}
\usepackage{bm}
\usepackage{gensymb}
\usepackage{indentfirst}
\usepackage{amsmath}
\usepackage{capt-of}
\usepackage{caption}
\usepackage{multirow}
\usepackage{rotating}
\usepackage{booktabs,caption}
\usepackage{tabularx}
\usepackage{color}
\usepackage{wasysym}
\usepackage{ctable}
\usepackage{tabularx,supertabular}
\usepackage{multirow}
\usepackage[utf8]{inputenc}
\usepackage[flushleft]{threeparttable}
\graphicspath{Figures} 
\begin{document}

\preprint{APS/123-QED}

\title{Theoretical determination of the ionization potentials of ScF, YF, LaF and AcF}

\author{A. A. Kyuberis}
\affiliation{Van Swinderen Institute for Particle Physics and Gravity, University of Groningen, Nijenborgh 3, 9747AG Groningen, The Netherlands}
\email{a.kiuberis@rug.nl}
\author{L. F. Pa\v{s}teka}
\affiliation{Van Swinderen Institute for Particle Physics and Gravity, University of Groningen, Nijenborgh 3, 9747AG Groningen, The Netherlands}
\affiliation{Department of Physical and Theoretical Chemistry, Faculty of Natural Sciences, Comenius University, Ilkovičova 6, 84215 Bratislava, Slovakia}
\author{E. Eliav}
\affiliation{School of Chemistry, Tel Aviv University, 6997801 Tel Aviv, Israel}
\author{A. Sunaga}
\affiliation{ELTE, E\"otv\"os Lor\'and University, Institute of Chemistry, P\'azm\'any P\'eter s\'et\'any 1/A 1117 Budapest, Hungary}
\author{M. Au}
\affiliation{Systems Department, CERN, CH-1211 Geneva 23, Switzerland}
\affiliation{Department of Chemistry, Johannes Gutenberg-Universit\"{a}t Mainz, 55099 Mainz, Germany}
\author{S. G. Wilkins}
\affiliation{Massachusetts Institute of Technology, Cambridge, MA 02139, USA}
\author{A. Borschevsky}
\affiliation{Van Swinderen Institute for Particle Physics and Gravity, University of Groningen, Nijenborgh 3, 9747AG Groningen, The Netherlands}

\date{\today}

\begin{abstract}
We present a comprehensive theoretical study of the ionization potentials of the $M$F ($M$ = Sc, Y, La, Ac) molecules using the state-of-the-art relativistic coupled cluster approach with single, double, and perturbative triple excitations (CCSD(T)). We have further corrected our results for higher-order excitations (up to full triples), the QED self-energy and vacuum-polarization contributions. We have extensively investigated the effect of the various computational parameters on the calculated ionization potentials, allowing us to assign realistic uncertainties to our predictions. 
\end{abstract}

\maketitle

\section{Introduction}

Ionization potentials (IPs) serve as fundamental descriptors in understanding the electronic structure and chemical behavior of atoms and molecules and are often necessary for guiding various experiments. However, precise measurements of molecular IPs can be a challenging task, especially for heavy molecules containing radioactive nuclei. Recently, radioactive elements have come into focus as an excellent source of new insights into symmetry-violating properties, due to their high atomic number and the nuclear octupole deformation of some of their isotopes~\cite{isaev2020lasercooled,arrowsmithkron2023opportunities,wilkins2024ionizationpotentialradiummonofluoride}. Molecules containing short-lived radioactive nuclei are in particular uniquely positioned to enable a wide range of scientific discoveries in the areas of fundamental symmetries \cite{PhysRevLett.116.161601}, astrophysics~\cite{Kamiński2018} and nuclear structure \cite{10.1063/5.0024103,GarciaRuiz2020c,SCHMIDT2018324}.
A series of experiments were recently conducted on the RaF molecule, where its ionization potential and excited state energies and lifetimes were accurately measured and analyzed~\cite{udr21,AthanasakisKaklamanakis2023,GarciaRuiz2020c,Garcia_Ruiz_2020,WilPer24}. The first study of RaF isotopologues demonstrated that even low-resolution molecular laser spectroscopy can allow the extraction of nuclear charge radii with a competitive systematic uncertainty, currently limited by the molecular calculations~\cite{udr21}. 



Theoretical calculations for nuclei with strong octupole deformation show that $^{227}$Ac should possess an exceptionally large nuclear Schiff moment, predicted to be the largest among all nuclei considered so far~\cite{PhysRevA.101.042504,PhysRevC.101.015502}. Additionally, the Schiff moment of $^{225}$Ac was also predicted to be remarkably large, 3 times that of $^{225}$Ra~\cite{PhysRevC.110.015501,PhysRevA.92.022502}. Therefore, the long-lived actinium isotopes are compelling candidates for the first successful measurement of the Schiff moment. Performing such a measurement in a molecule rather than in an atom carries many advantages, such as the simpler electronic structure, which in turn simplifies the interpretation of the experiment and the extraction of the Schiff moment, and the fact that the electronic structure actually can lead to further enhancement of the effect of interest \cite{arrowsmithkron2023opportunities}. A promising molecular candidate containing the Ac atom is AcF \cite{SkrMos20}. 
Planning molecular precision measurements requires accurate knowledge of the electronic structure and fundamental properties of the candidate molecule, including the ionization potential \cite{KakWil21}. Thus, the first step towards precision measurements of the Schiff moment in this promising system is its spectroscopic investigations.

Spectroscopic studies of heavy or unstable atoms and molecules containing such isotopes are inherently challenging and greatly benefit from strong and reliable theoretical support \cite{Arrowsmith-Kron_2024}. It is therefore the aim of this work to provide accurate predictions of the ionization potential of AcF, along with those of its lighter homologue LaF, and two molecules with a similar electronic structure, i.e. YF and ScF. We use the state-of-the-art relativistic coupled cluster method; to achieve optimal accuracy, we use complete basis set limit extrapolations higher-order relativistic and correlation corrections. We also carry out an extensive computational investigation that allows us to assign robust uncertainties to the predicted IPs.

The measured ionization potentials of YF, ScF, and LaF are reported in Ref. \cite{ZmbMar68}, where the ionization potentials, bond energies, atomization energies, heats of sublimation, and heats of formation were obtained through a mass spectrometric study of the mono-, di-, and trifluorides of scandium, yttrium, lanthanum, and other rare-earth metals. The authors set the accuracy of the measured ionization potentials at about $\pm$ 0.3 eV. 
So far, no accurate measurements of the IPs of any of the systems discussed here were performed using Rydberg spectroscopy or photoionization threshold as was done, for example, for CaF, BaF, and RaF~\cite{Berg1993,Jakubek1995,wilkins2024ionizationpotentialradiummonofluoride}. 

Other properties of these molecules were investigated, both experimentally and theoretically. For ScF, both experimental~\cite{KALEDIN1995569,LebEff94,McLWel66} and theoretical~\cite{SolMuk12} studies are available, focusing on the spectroscopic constants, such as $R_e$, $\omega_e$, and $\omega_{e}\chi_{e}$ for the different electronic states. Similarly, a number of experiments were performed on LaF~\cite{Kaledin:94,SchLin83,Ass22,RUBINOFF2003169,BerEff00,HilLau95}, also focusing on the spectroscopic constants of its various states in addition to a theoretical~\cite{10.1063/1.1493769} study. 
For YF, even less experimental data is available~\cite{HamLes82,SHENYAVSKAYA197623}. 
In Ref. \cite{HamLes82} the absorption and emission spectra for YF in low temperature matrices are reported. Upon argon laser excitation, YF produces an extremely intense, well-resolved fluorescence spectrum from higher vibrational levels of the $C^{1}{\Sigma}^{+}$ state. The available theoretical work \cite{ABDULAL2005183} presents investigations of the spectroscopic constants of the lowest molecular states of YF, along with the  transition energies. There, the lowest molecular states of YF were investigated via a combination of complete active space self consistent field approach (CASSCF) and multi reference configuration interaction method (MRCI).
Finally, the spectroscopic and radiative properties of AcF were recently investigated in depth using the relativistic coupled cluster approach \cite{SkrMos20,SkrOle23}.

\section{Method and and computational Details}

The calculations were carried out using the official release of the DIRAC program~\cite{DIRAC23}, which allows us to use relativistic methods, in particular the traditional 4-component Dirac--Coulomb (DC) Hamiltonian. The single-reference coupled cluster approach with single, double (CCSD), and perturbative triple excitations (CCSD(T)) was used in the baseline calculations. 
To take full advantage of this state-of-the-art approach, we used relativistic Dyall basis sets~\cite{dyall3,dyall2}. These basis sets are available in cardinalities $n$ = 2, 3, 4. Furthermore, one can employ either the “valence” (dyall.v$n$z), the “core-valence” (dyall.cv$n$z), or the “all-electron" (dyall.ae$n$z) variants of the basis sets of the same cardinality; the latter two include additional core-valence- and core-correlating functions, respectively. Finally, extra layers of diffuse functions were  added, designated by $k$-aug-(v/cv/ae)$n$z, with $k$ = s (single), d (double), etc. The presence of diffuse functions improves the description of the bond region and, hence, of the valence properties like ionization potentials. 

The adiabatic IPs were derived from the molecular energies calculated at the respective equilibrium bond lengths of the neutral molecules and molecular ions. For this purpose, the potential energy curves of the four molecules and their ions were calculated. 
The baseline results for the IPs and the rest of the molecular properties were obtained from a complete basis set limit (CBSL) extrapolation of the potential energy curves based on the the s-aug-cv$n$z ($n=2,3,4$) basis sets and carried out following the scheme of Helgaker~\textit{et al.}~\cite{cbs} (H-CBSL). 
In our recent work~\cite{kyuPas23}, we found that results based on different CBS schemes (schemes of Helgaker~\textit{et al.}~\cite{cbs} (H-CBSL), Martin~\cite{cbs1} and of Lesiuk and Jeziorski~\cite{cbs2}) are consistent to within 1 meV, confirming the convergence of the calculated IPs with respect to the basis set cardinality. Based on this, we use the H-CBSL scheme in this work. In these calculations, 26, 36, 36, and 50 electrons were correlated for YF, ScF, LaF, and AcF, respectively, corresponding to an active space cutoff of --20~a.u., while the virtual space cutoff was set at 50~a.u. 
With the aim of achieving the highest possible accuracy, we took into account additional higher-order corrections, such as higher CC excitations and relativistic contributions beyond the DC Hamiltonian. 

\section{Results}

Table~\ref{constants} contains the calculated CCSD(T) spectroscopic constants of the ground states of the neutral molecules and molecular ions, compared to earlier theoretical predictions and experimental values, where available. All four molecules are predicted to have $ X~^1 \Sigma $ ground state, while the corresponding molecular ions have $ X~^2 \Sigma_{1/2} $ ground state. Our results are in a good agreement with previous calculations and experimental values.

  \begin{table}[h]
        \centering
        \caption{Spectroscopic constants of the $M$F ($M$ = Sc, Y, La, Ac) molecules and the corresponding $M$F$^+$ ions, calculated in this work (TW) and compared to previously determined theoretical and experimental values.}\label{constants}
\begin{tabular}{p{15mm}lllc}
        \hline
$R_e$(\AA)&$\omega_e$ (cm$^{-1}$)&$\omega_{e}\chi_{e}$ (cm$^{-1}$)&Ref.&Method\\
              \hline
\noalign{\vspace{4pt}}  
\multicolumn{5}{c}{ScF}              \\
\noalign{\vspace{4pt}}  
1.786&741&3.51&TW &CCSD(T)\\
1.787&737&3.57&~\cite{SolMuk12}&CCSDT\\
1.787&736&3.80&~\cite{McLWel66}&Exp.\\
1.787&735&3.77&~\cite{LebEff94}&Exp.\\
--&735&3.60&~\cite{HilLau95}&Exp.\\
\noalign{\vspace{4pt}}   
\multicolumn{5}{c}{ScF$^{+}$}\\
  \noalign{\vspace{4pt}}
1.749&775&4.83&TW&CCSD(T)\\
\hline
\multicolumn{5}{c}{YF} \\
1.924&638&2.76&TW&CCSD(T)\\
1.928&644&--&~\cite{ABDULAL2005183}&MRCI+SO\\
--&636&2.50&~\cite{HamLes82}&Exp.\\
--&635&2.50&~\cite{HilLau95}&Exp.\\
\noalign{\vspace{4pt}}    
\multicolumn{5}{c}{YF$^{+}$} \\
  \noalign{\vspace{4pt}}
1.872&717&3.41&TW&CCSD(T)\\
1.876&--&--&~\cite{SHENYAVSKAYA197623}&Exp.\\
\hline
\multicolumn{5}{c}{LaF} \\
2.023&578&2.43&TW&CCSD(T)\\
2.067&598&--&~\cite{10.1063/1.1493769}&CI$^{b}$\\
2.057&583&--&~\cite{10.1063/1.1493769}&CI+Q$^{c}$\\
2.035&574&2.4&~\cite{Ass22}&MRCI$^{a}$\\
--&570&2.3&~\cite{HilLau95}&Exp.\\
2.023&574.948(25)&2.11302(14)&~\cite{RUBINOFF2003169}&Exp.\\
      \noalign{\vspace{4pt}}
\multicolumn{5}{c}{LaF$^{+}$} \\
  \noalign{\vspace{4pt}}
1.972&635&2.08&TW&CCSD(T)\\
              \hline
\multicolumn{5}{c}{AcF} \\
2.103&531&3.87&TW&CCSD(T)\\
2.110&541&--&~\cite{SkrOle23}&IH-FSCCSD$^{d}$\\
      \noalign{\vspace{4pt}}
\multicolumn{5}{c}{AcF$^{+}$} \\
  \noalign{\vspace{4pt}}
2.047&602&1.75&TW&CCSD(T)\\
\hline
        \end{tabular}
        \begin{tablenotes}
        \footnotesize
        \item $^{a}$ Multi Reference Configuration Interaction (MRCI).
        \item $^{b}$ Configuration interaction (CI).
        \item $^{c}$ Configuration interaction with Pople/Davidson correction (CI+Q).
        \item $^{d}$ Intermediate Hamiltonian Fock Space Coupled Cluster (IH-FSCCSD).   
        \end{tablenotes}
    \end{table} 

\begin{table}[h]
\setlength{\tabcolsep}{5pt}
        \centering
        \caption{Adiabatic IPs (eV) of the $M$F ($M$ = Sc, Y, La, Ac) molecules calculated using basis sets of differing quality. The H-CBSL extrapolated value is labeled $\infty$z.}\label{cbs-sch}
        \begin{tabular}{lccccc}
        \hline
              Basis set &ScF&YF&LaF&AcF \\
              \hline
              dyall.v4z&7.0643&6.6755&5.9798&6.0438\\
              dyall.cv4z&7.0605&6.6817&5.9299&6.0457\\
              dyall.ae4z&7.0605&6.6817&5.9223&6.0475\\
              s-aug-dyall.cv2z&7.1420&6.6950&6.0063&6.1149\\
              s-aug-dyall.cv3z&7.0825&6.6860&5.9510&6.0598\\
              s-aug-dyall.cv4z&7.0657&6.6848&5.9326&6.0468\\
              d-aug-dyall.cv4z&7.0660&6.6852&5.9332&6.0471\\
              \textbf{s-aug-dyall.cv$\infty$z} &\textbf{7.0550}&\textbf{6.6876}&\textbf{5.9294}&\textbf{6.0466}\\
        \hline
        \end{tabular}
    \end{table} 

Table~\ref{cbs-sch} presents the IP values obtained using different basis sets demonstrating the convergent behavior of the ionization potentials as progressively larger and more comprehensive basis sets are used. The same active space cutoff of --20~a.u. to 50~a.u. was used in these calculations as those presented in Table \ref{constants}. 
Switching from the valence (dyall.v4z) to the core valence (dyall.cv4z) basis sets has an effect of a few meV on the calculated IPs of all the systems except LaF, where the difference is much more pronounced, at 50 meV. This behavior of Dyall basis sets for La atom is under investigation. 

There is no further significant change in the IPs of ScF, YF, and AcF when going to the dyall.ae4z basis; for LaF the effect is around 8 meV. We thus use the core-valence basis set family for our final results. 
Based on the results in Table~\ref{cbs-sch}, we also conclude that going from single to double augmentation does not significantly affect the calculated IPs and we use the singly augmented cvXz values for the extrapolation to the complete basis set limit (shown in Table \ref{cbs-sch} in bold font). We then proceed to correct these CBS values for the various effects that were neglected in the baseline calculation.

To investigate the effect of the limited active space used, we calculated the difference between the results obtained correlating electrons occupying orbitals with energies above --20 a.u. and a virtual space cutoff of 50 a.u. (as the results shown in Table \ref{cbs-sch} were obtained) and those obtained correlating all electrons (30, 48, 66 and 98, for ScF, YF, LaF and AcF, respectively) with a corresponding virtual space cutoff of 4000 a.u. for all the systems. In order to capture the full active space effect and to account for the inner-core correlations, the all-electron quality basis sets were used in the latter calculations. For these calculations, we used the dyall.cv4z and dyall.ae4z basis sets. The results are shown in Table~\ref{active_space}, and the differences between the small active space dyall.cv4z values and the large active space dyall.ae4z values were used to correct the CBS-extrapolated results (line 3 in Table~\ref{Corrections}).

\begin{table}[h]
\setlength{\tabcolsep}{5pt}
       \centering
        \caption{Calculated adiabatic IPs (eV) of $M$F ($M$ = Sc, Y, La, Ac) calculated using different active space sizes; small active space: --20~ to ~50~a.u.; large active space: all electrons correlated and virtual cut-off of 4000~a.u.}\label{active_space}
        \begin{tabular}{lccccc}
        \hline
              Basis set & active space&ScF&YF&LaF&AcF\\
              \hline
          dyall.cv4z&small&7.0605&6.6817&5.9299&6.0457\\
          dyall.cv4z&large&7.0646&6.6813&5.9313&6.0476\\
          dyall.ae4z&large&7.0646&6.6814&5.9231&6.0516\\
        \hline
        \end{tabular}
    \end{table} 

 \begin{table}
\setlength{\tabcolsep}{5pt}
        \centering
        \caption{Calculated adiabatic IPs (eV) of $M$F ($M$ = Sc, Y, La, Ac), including higher-order corrections. The DC-CCSD and DC-CCSD(T) results are presented at the CBS limit based on the s-aug-dyall.cv$n$z basis sets, correlating 26, 36, 36, and 50 electrons, respectively, and including virtual orbitals up to 50 a.u.}\label{Corrections}
        \begin{tabular}{lccccc}
        \hline
    Method & ScF &YF &LaF&AcF \\
     \hline 
    DC-CCSD&\phantom{--}6.9720&\phantom{--}6.5891&\phantom{--}5.7973&\phantom{--}5.9210\\
    DC-CCSD(T) &\phantom{--}7.0550&\phantom{--}6.6876&\phantom{--}5.9294&\phantom{--}6.0466\\
    +$\Delta$active space &\phantom{--}0.0041&-0.0003&-0.0068&\phantom{--}0.0059\\
    +$\Delta$augmentation&\phantom{--}0.0003&\phantom{--}0.0005&\phantom{--}0.0006&\phantom{--}0.0003\\
    +$\Delta$T &\phantom{--}0.0023 &\phantom{--}0.0017&\phantom{--}0.0017&\phantom{--}0.0001\\
    +$\Delta$Breit &\phantom{--}0.0069&\phantom{--}0.0069&\phantom{--}0.0070&--0.0030\\
    +$\Delta$QED &--0.0012&--0.0020&--0.0030&--0.0062\\
        \hline
       Final result &\phantom{--}7.0674&\phantom{--}6.6943&\phantom{--}5.9289&\phantom{--}6.0438\\
             \hline
        \end{tabular}
    \end{table} 

We corrected our results for the effect of possible insufficient amount of diffuse functions by taking the difference between the s-aug-dyall.cv4z and the d-aug-dyall.cv4z values (Table~\ref{cbs-sch} and line 4 in Table \ref{Corrections}). 

We evaluated the effect of the residual triple excitations (beyond (T)) by comparing the IPs calculated at the CCSDT and CCSD(T) levels of theory using the MRCC code~\cite{MRCC1}. These calculations were performed using the dyall.v3z basis sets, correlating 9, 9, 18, and 18 electrons for ScF, YF, LaF, and AcF, respectively, and with a virtual space cutoff set at 5~a.u. As the effect was very small (about 2~meV in most cases, see row 5 in Table \ref{Corrections}), we did not consider excitations beyond triples.

There is presently no straightforward four-component 
implementation for calculating the Breit effect in molecules
on the correlation level. We thus take advantage of the fact that the electronic structure of the $M$F molecules is similar to that of the $M^+$ ions and calculated the effect of the Breit contribution on the IPs of $M^+$ ions. The Breit calculations were performed within the Fock-space coupled cluster approach (FSCC), using the Tel Aviv atomic computational package~\cite{TRAFS-3C}; this effect contributes a few meV to the final IPs (row 6 in Table \ref{Corrections}).

The QED corrections were calculated using the development version of the DIRAC program package~\cite{SunSalSau22}. The Uehling potential~\cite{Ueh35} was employed for the vacuum polarization, and the effective potential of Flambaum and Ginges for the electron self-energy~\cite{FlaGin05}. The results are shown in row 7 in Table \ref{Corrections}, and they lower the IP by a few meV for all molecules considered here.  

All the corrections were calculated at the equilibrium geometries of the neutral $M$F molecules and of their ions and added to the baseline IPs obtained at the CBS limit (Table~\ref{cbs-sch}) to obtain the final recommended values of IPs, presented in the bottom row of Table \ref{Corrections}. The active space and relativistic corrections have the largest impact on the calculated IPs; with the latter being particularly prominent for the heavier systems. The total effect of the higher-order corrections ranges from 12 meV for ScF to just 1-2 meV for the heavier LaF and AcF, where the increased magnitude of the QED corrections mostly cancels out the other effects. 

\section{Uncertainty estimates}

Reliable estimates of the uncertainty of theoretical predictions are crucial for supporting and interpreting experiments on the molecules in question. These uncertainty estimates can be obtained from extensive computational investigations; we have successfully employed such procedures in the past for various atomic and molecular properties~\cite{prop1,prop2,prop3,LeiKarGuo20,kyuPas23}. 

The three main sources of uncertainty in our calculations are the incompleteness of the employed basis set, approximations in the treatment of the electron correlation and missing relativistic effects. As we are considering higher-order effects, we assume these sources of error to be largely independent, and hence, we can treat them separately. 
In the following, each of these contributions will be presented and evaluated separately; the individual contributions to the total uncertainty are given in Table~\ref{uncertainty-all}. 

\paragraph{Basis set.}\label{bst}
To evaluate the basis set incompleteness error, we consider the basis set cardinality and the convergence in terms of core-correlating and diffuse functions. 
The final results were obtained at the CBS level. We evaluate the cardinality incompleteness error as half of the difference between CBS and s-aug-dyall.cv4z basis set results. In order to evaluate the uncertainty due to the missing inner-core correlating functions, we estimate the difference between the calculation with the all-electron quality basis set (dyall.ae4z) and standard dyall.cv4z basis set, while correlating all the electrons with a corresponding virtual space cutoff of 4000 a.u..
The remaining uncertainty due to the possible lack of additional diffuse functions (beyond double augmentation) is evaluated as the difference between the results obtained using the singly- and doubly- augmented dyall.v4z basis sets. 
 
\paragraph{Electron correlation.}\label{ctt}
We consider separately the effect of using a limited active space (virtual space cutoff) and the effect of excitations beyond perturbative triples.
To account for the limited active space, we take the difference between results obtained with a virtual cutoff of 50~a.u. and 4000~a.u. at the dyall.cv4z basis set level correlating 26, 36, 36 and 50 electrons for ScF, YF, LaF, and AcF, respectively.
studies of IPs of heavy atoms \cite{Pasteka2017,Guo2021,Guo_2022}, the size of the contributions beyond triples were at the level of single meV. To estimate the uncertainty due to the neglected higher excitation contributions beyond CCSDT, we use a conservative 10$\%$ fraction of the triples contribution. This is a somewhat conservative estimate, as the higher-order corrections typically exhibit alternating signs leading to partial cancellations \cite{RUDEN200362,10.1063/1.1780155}.

\paragraph{Relativistic effects.}\label{rtt}



The various effects contributing to the uncertainty are given in the Table~\ref{uncertainty-all} and the total uncertainty is obtained by combining all the above terms and assuming them to be independent. We find a rather consistent uncertainty of 10--15 meV for all the systems considered here. 

    \begin{table}[h]
    \setlength{\tabcolsep}{5pt}
        \centering
        \caption{Main sources of uncertainty of the calculated IPs (meV) of $M$F ($M$ = Sc, Y, La, Ac).} 
        \label{uncertainty-all}
        \begin{tabular}{llcccc}
        \hline
              Category & Error source& ScF&YF& LaF &AcF\\
              \hline
        Basis set & Cardinality&5.38&1.46&1.61&0.07 \\
                  & augmentation&0.31&0.46&0.62&0.32\\
                  & core correlation&0.02&0.01&8.17&4.05\\
        Correlation& virtual space &3.10 &0.36&0.17&0.88\\ 
                   & higher excitations&8.76&10.19&13.54&12.59\\
        Relativity & QED & 0.19&0.56&1.25&4.01\\
        \hline
        Uncertainty&&10.74&10.32&15.96&13.86\\
        \hline
        \end{tabular}
    \end{table}

\section{Final values}

The final recommended adiabatic IPs of ScF, YF, LaF, and AcF, along with the corresponding uncertainties, are presented in Table~\ref{IPfinal}.
To make a meaningful comparison with experiment, we also present the vibrationally corrected adiabatic IP$_{0-0}$ corresponding to the 0--0 transition and explicitly accounts for the zero point vibrational energy (ZPVE), approximated as:  

\begin{gather}
\notag     \mathrm{IP}_{0-0} =(E^{M\mathrm{F^{+}}}+\frac{1}{2} \omega_e^{M\mathrm{F^{+}}}-\frac{1}{4} {\omega_{e}\chi_{e}}^{M\mathrm{F^{+}}})-\\
 (E^{M\mathrm{F}}+\frac{1}{2} \omega_e^{M\mathrm{F}}-\frac{1}{4}{\omega_{e}\chi_{e}}^{M\mathrm{F}}),
\end{gather}
where $E^{M\mathrm{F{^+}}}$ and $E^{M\mathrm{F}}$ are the energies of the ion and the neutral systems ($M$ = Sc, Y, La, Ac) at the corresponding equilibrium bond lengths,  
$\omega_e^{M\mathrm{F^{+}}}$ is the vibrational frequency of $M$F$^{+}$ and $\omega_e^{M\mathrm{F}}$ is the vibrational frequency of $M$F (Table~\ref{constants}).
Vertical IPs are calculated as the energy difference between the ionized and neutral states with the molecular geometry remaining fixed at the $R_{e}$ of the neutral molecule.

    \begin{table}[h]
        \centering
        \caption{Recommended theoretical IPs (eV) of $M$F ($M$ = Sc, Y, La, Ac) with uncertainties compared to available experimental and theoretical data.}\label{IPfinal}
        \begin{tabular}{lllll}
        \hline
        IP & ScF&YF&LaF&AcF  \\
        \hline 
        Vertical&7.071&6.728&5.956&6.077\\
        Adiabatic &7.067(11)&6.694(10)&5.929(16)&6.044(14)\\  
        Adiab.+ZPVE&7.070(11)&6.699(10)&5.932(16)&6.048(14)\\
        \hline
        Theory~\cite{SkrOle23}&--&--&--&6.058(16)\\
        Experiment~\cite{ZmbMar68} &6.5(3)&6.3(3) &6.3(3)&-- \\
        \hline
        \end{tabular}
    \end{table} 

Experimental IPs of ScF, YF and LaF were extracted from an extensive mass spectrometric study of the sublimation rates for solid rare-earth trifluorides and of the stabilities of the various gaseous fluoride species which are present in the gaseous mixture at equilibrium~\cite{ZmbMar68}. These measurements, performed using the Knudsen cell technique, have an estimated uncertainty of $\pm 0.3$ eV, which is large compared to the precision of the current theoretical predictions and achievable accuracy of the experimental state-of-the-art techniques. Moreover, the calibration of these experiments relied on an outdated IP value for mercury, introducing systematic errors that further compromise the reliability of the experimental data. Hence, we cannot expect good agreement between the calculated ionization potentials of ScF, YF, and LaF and the experimental data from Ref. \cite{ZmbMar68}. Indeed, we find the current predictions to be higher by 0.3--0.5 eV than the measured values, motivating reevaluation of the latter. The significant uncertainties and the outdated nature of the experimental data for ScF, YF, and LaF highlight the need for modern techniques to measure IPs more precisely. Methods such as ionization threshold spectroscopy \cite{WilPer24} or Rydberg series analysis \cite{Rothe2013} could provide more reliable benchmarks for comparison.

For AcF, no experimental IP measurements are available. We can however compare the present result to recent calculations performed using a very similar approach, namely relativistic coupled cluster \cite{SkrOle23}. The two results differ by about 10 meV, which is within the combined theoretical uncertainties. 
The main factors contributing to the discrepancy of the two values are the inclusion of perturbative quadruple excitation in Ref. \cite{SkrOle23} and a different choice of the employed basis sets. 

\begin{figure}[h]
    \centering
    \includegraphics[width=250pt]{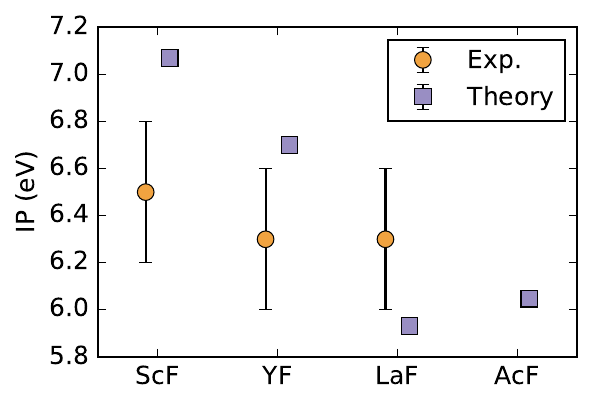}
    \caption{Comparison between predicted theoretical and measured values \cite{ZmbMar68} for the IP of Group III monofluorides. The error bars for the theoretical predictions are smaller than the markers.}
    \label{fig:comparsion}
\end{figure}

The observed trend in theoretical IPs as a function of the atomic number $Z$ reveals a zig-zag pattern (Fig. \ref{fig:comparsion}), which can be attributed to relativistic effects. These effects alter the energies of the orbitals from which the electron is removed \cite{pyy12}. Orbital analysis reveals that the electron removed during ionization has a predominant $s$-character across all the molecules considered here.
The strong relativistic contraction and stabilization of the $s$-orbital in the heaviest molecule AcF leads to a reversal of the trend of reduction of the IP with the increase of atomic number observed up to LaF. Ongoing experimental campaigns involving radioactive molecules, such as those conducted at CERN-ISOLDE, may offer opportunities for validating the theoretical predictions reported here.While so far there is no existing measurement of the IP of AcF, the recent experimental campaign on the production of actinium fluorides at CERN-ISOLDE made use of a FEBIAD-type ion source \cite{PenCat10}, which ionizes and dissociates molecules using accelerated electrons, similarlto the Knudsen cell technique used to measure appearance potentials of the lanthanide fluorides. in Ref. \cite{ZmbMar68} The AcF molecule, conclusively identified using combined mass spectrometry and decay spectroscopy, was observed with electron energies above 25\,eV. However, the use of the ion source as a technique for determining appearance potentials is limited by the wide distribution of electron energies in the device, and as such can only suggest an upper bound, necessitating mdedicated easurements performed using accurate techniques for the confirmation of the current predictions.

\section{Conclusions}
We performed calculations of the IPs of ScF, YF, LaF, and AcF using the 4-component CCSD(T) method and employing large uncontracted basis sets with extrapolation to the complete basis set limit.
To further improve the accuracy, the Breit, QED and higher-order excitation corrections were added a posteriori to the CCSD(T) values. An extensive analysis of the effect of the various computational parameters allowed us to assign realistic uncertainties of 10--15 meV to our predictions.
For all the molecules studied in this work, the theoretical evaluation of the IP is expected to be more accurate than the available experimental data on the same species. The results reported in this work will contribute towards future experimental campaigns towards measuring the IPs of these molecules using advanced experimental methods.

\section{Acknowledgments}

We thank the Center for Information Technology at the University of Groningen for their support and for providing access to the Peregrine high performance computing cluster and to the Hábrók high performance computing cluster. Authors thank Dr. Michail Athanasakis-Kaklamanakis for useful discussions and help with navigating through experimental data. The work of AB and AK was supported by the project \textit{High Sector Fock space coupled cluster method: benchmark accuracy across the periodic table} with project number Vi.Vidi.192.088 of the research programme Vidi which is financed by the Dutch Research Council (NWO).
AS thanks JSPS Overseas Challenge Program for Young Researchers, Grant No. 20188019.
LFP acknowledges the support from the Slovak Research and Development Agency (projects APVV-20-0098, APVV-20-0127) and from the Scientific Grant Agency of the Slovak Republic (project 1/0254/24).

\bibliography{apssamp}

\end{document}